\documentclass[aps,prl,twocolumn,superscriptaddress]{revtex4-1}
\usepackage{graphicx}
\usepackage{natbib}
\usepackage{color}
\usepackage{amsmath}
\usepackage{amssymb}
\usepackage{array}
\usepackage{wasysym}
\usepackage{bm}
\usepackage{textcomp}

\begin{document}

% Use the \preprint command to place your local institutional report
% number in the upper righthand corner of the title page in preprint mode.
% Multiple \preprint commands are allowed.
% Use the 'preprintnumbers' class option to override journal defaults
% to display numbers if necessary
%\preprint{}
%Title of paper
\title{Classical Spin Nematic Transition in LiGa$_{0.95}$In$_{0.05}$Cr$_4$O$_8$}

% repeat the \author .. \affiliation  etc. as needed
% \email, \thanks, \homepage, \altaffiliation all apply to the current
% author. Explanatory text should go in the []'s, actual e-mail
% address or url should go in the {}'s for \email and \homepage.
% Please use the appropriate macro foreach each type of information

% \affiliation command applies to all authors since the last
% \affiliation command. The \affiliation command should follow the
% other information
% \affiliation can be followed by \email, \homepage, \thanks as well.
\author{R. Wawrzy\'{n}czak}
\email[Email address:~]{wawrzynczak@ill.eu}
\affiliation{Institut Laue-Langevin, 6 rue Jules Horowitz, 38042 Grenoble, France}

\author{Y. Tanaka}
\email[Email address:~]{you@issp.u-tokyo.ac.jp}
\affiliation{Institute for Solid State Physics, University of Tokyo, 5-1-5 Kashiwanoha, Kashiwa, Chiba 277-8581, Japan}

\author{M. Yoshida}
\affiliation{Institute for Solid State Physics, University of Tokyo, 5-1-5 Kashiwanoha, Kashiwa, Chiba 277-8581, Japan}

\author{Y.~Okamoto}
\affiliation{Department of Applied Physics, Nagoya University, Furo-cho, Chikusa-ku, Nagoya 464-8603, Japan}

\author{P.~Manuel}
\affiliation{ISIS Neutron and Muon Source, Science and Technology Facilities Council, Didcot, OX11 0QX, United Kingdom}

\author{N.~Casati}
\affiliation{Swiss Light Source, Paul Scherrer Institute, 5232 Viligen PSI, Switzerland}

\author{Z.~Hiroi}
\affiliation{Institute for Solid State Physics, University of Tokyo, 5-1-5 Kashiwanoha, Kashiwa, Chiba 277-8581, Japan}

\author{M.~Takigawa}
\affiliation{Institute for Solid State Physics, University of Tokyo, 5-1-5 Kashiwanoha, Kashiwa, Chiba 277-8581, Japan}

\author{G.~J.~Nilsen}
\email[Email address:~]{goran.nilsen@stfc.ac.uk}
\affiliation{ISIS Neutron and Muon Source, Science and Technology Facilities Council, Didcot, OX11 0QX, United Kingdom}

%Collaboration name if desired (requires use of superscriptaddress
%option in \documentclass). \noaffiliation is required (may also be
%used with the \author command).
%\collaboration can be followed by \email, \homepage, \thanks as well.
%\collaboration{}
%\noaffiliationgreen

\date{\today}

\begin{abstract}We present the results of a combined $^7$Li NMR and diffraction study on LiGa$_{0.95}$In$_{0.05}$Cr$_4$O$_8$, a member of LiGa$_{1-x}$In$_{x}$Cr$_4$O$_8$ ``breathing'' pyrochlore family. Via specific heat and NMR measurements, we find that the complex sequence of first-order transitions observed for $x=0$ is replaced by a single, apparently second-order transition at $T_f=11$ K. Neutron and X-ray diffraction rule out both structural symmetry lowering and magnetic long-range order as the origin of this transition. Instead, reverse Monte Carlo fitting of the magnetic diffuse scattering indicates that the low temperature phase may be described as a collinear spin nematic state, characterized by a quadrupolar order parameter. This state also shows signs of short range order between collinear spin arrangements on tetrahedra, revealed by mapping the reverse Monte Carlo spin configurations onto a three-state color model describing the manifold of nematic states. 
\end{abstract}

\maketitle

	Spinel materials, $AB_2X_4$, host a variety of interesting magnetic phenomena, including spin-orbital liquid states (FeSc$_2$S$_4$) \cite{fritsch04}, skyrmion lattices (GaV$_4$S$_8$) \cite{kezsmarki15}, and magneto-structural transitions \cite{lee00,ji09,bordacs09,nilsen15}. Many of these originate from the $B$ site, which forms a frustrated pyrochlore lattice of corner-sharing tetrahedra. When the large spin degeneracy caused by the frustration is combined with strong magneto-elastic coupling, a typical feature of spinels, several possible magneto-structurally ordered and disordered states arise. For example, in the chromate spinel oxides, $A$Cr$_2$O$_4$, collinear, coplanar, and helical magnetic structures (and their accompanying structural distortions) may all be realized by varying the cation on the $A$ site \cite{nilsen15,lee00,ji09,chung05}. Although the low temperature behaviour of the chromates is complex, it is surprisingly well captured by the bilinear-biquadratic model \cite{tchernyshyov02,shannon10}, the Hamiltonian of which is $\mathcal{H}=J\sum_{i,j}{\vec{S}_i\cdot\vec{S}_j}+b\sum_{i,j}{(\vec{S}_i\cdot\vec{S}_j)^2}+\mathcal{P}$, where $\vec{S}_{i,j}$ are classical Heisenberg spins and $J$ is the nearest neighbour exchange. The second and third terms represent the magneto-elastic coupling, which assumes a biquadratic form if only local distortions are considered, and perturbative terms such as further neighbor or anisotropic couplings. These act on the degenerate manifold of states $\sum_{i\in\textrm{tet}}{\vec{S}_i=0}$ (where the sum is over tetrahedra) generated by the Heisenberg term, in turn selecting either collinear ($b<0$) or coplanar ($b>0$) configurations \cite{tchernyshyov02}, then breaking the remaining degeneracy and establishing magnetic order ($\mathcal{P}\neq 0$) \cite{chern06,chern08}.
	
	Two recent additions to the chromate spinel family are the so-called ``breathing'' pyrochlore systems, $AA^\prime$Cr$_4$O$_8$, where $A$=Li$^+$ and $A^\prime$=Ga$^{3+}$,In$^{3+}$ \cite{okamoto13}. Here, the alternation of the $A$ and $A^\prime$ cations on the $A$ site leads to an alternation in tetrahedron sizes and, hence, magnetic exchange constants, $J$ and $J^\prime$. This alternation is quantified by the ``breathing'' factor $B_f=J^\prime{}/J$, which is $\sim 0.6$ for $A^\prime$=Ga$^{3+}$ and $\sim 0.1$ for $A^\prime$=In$^{3+}$. The small $B_f$ notwithstanding, the phenomenology of the ``breathing'' pyrochlores at low temperature is similar to their undistorted cousins. In both $A^\prime=$Ga$^{3+}$ ($x=0$) and $A^\prime=$In$^{3+}$ ($x=1$), a sequence of two transitions, as observed in lightly doped MgCr$_2$O$_4$ \cite{kemei14}, lead to structural and magnetic phase separation (as in ZnCr$_2$O$_4$ \cite{kemei11}), while the excitation spectra show gapped ``molecular'' modes in the ordered states \cite{tomiyasu08,nilsen15}. For $x=0$, the upper first-order magneto-structural transition at $T_u\sim 20$~K results in phase-separation into cubic paramagnetic and tetragonal collinear phases \cite{magstr}. The cubic phase then undergoes another first-order transition into a second tetragonal phase at $T_l=13.8$~K, the structure of which has not yet been determined. Interestingly, this transition is preceded by a divergence in the $^7$Li NMR $1/T_1$, implying proximity to a tricritical point \cite{tanaka14}.
	
	Studies of the solid solutions LiGa$_{1-x}$In$_{x}$Cr$_4$O$_8$ indicate that $T_l$ is rapidly suppressed when $x$ is increased/decreased from $x=0/1$. Starting from the $x=0$ composition, sharp peaks in the magnetic susceptibility and specific heat persist until $x\sim 0.1$, beyond which they are replaced by features characteristic of a spin glass \cite{okamoto15}. This side of the phase diagram resembles those of both the undistorted chromate oxides \cite{ratcliff02} and Monte Carlo (MC) simulations of the bilinear-biquadratic model for $b<0$ including both disorder and further neighbor interactions $\mathcal{P}=J_{nnn}\sum_{i,j}\vec{S}_i\cdot\vec{S}_j$ \cite{shinaoka14}.

	In this letter we will show that the low temperature behaviour of the $x = 0.05$ composition, apparently well inside the magnetically ordered regime of the phase diagram, differs drastically from the $x=0$ composition. Instead of two first-order transitions and phase separation, we observe a single second-order transition using $^7$Li NMR and specific heat. Remarkably, this transition neither corresponds to magnetic long range order nor a structural transition. Rather, the magnetic diffuse neutron scattering implies that it shares features with both the nematic transition predicted for the bilinear biquadratic model on the pyrochlore lattice, as well as the partial ordering transition expected for the pyrochlore antiferromagnet with further neighbor interactions \cite{chern08}. Furthermore, the transition is shown to coincide with spin freezing, drawing parallels to other frustrated materials, like Y$_2$Mo$_2$O$_7$ \cite{silverstein14,gardner99}.
	
	The powder samples of LiGa$_{0.95}$In$_{0.05}$Cr$_4$O$_8$ were prepared via the solid-state reaction method in Ref. \cite{okamoto13}, using $^7$Li enriched starting materials to reduce neutron absorption. The specific heat was measured by the relaxation method in a Physical Property Measurement System (Quantum Design). The $^{7}$Li-NMR measurements were carried out in a magnetic field of $2$~T, and NMR spectra were obtained by Fourier transforming the spin-echo signal.  At low temperature, the spectra were constructed by summing the Fourier transformed spin-echo signals measured at equally spaced frequencies. The nuclear spin-lattice relaxation rate 1/$T_1$ was determined by the inversion-recovery method at the spectral peak \cite{suppmat}.

	The samples were further characterised by powder synchrotron X-ray diffraction (SXRD) on the MS-X04SA beamline at the Swiss Light Source, and powder neutron time-of-flight diffraction (ND) on the WISH instrument at the ISIS facility. For the former, several patterns were measured in the temperature range $6-20$~K using a photon energy of $22$~keV ($\lambda=0.564$ \AA). The ND was measured at several temperatures between $1.5$~K and $30$~K. Structural refinements were carried out on data from all 10 detector banks on WISH as well as the SXRD data [Fig. \ref{FIG2}(a)] using the FULLPROF software \cite{fullprof}. The magnetic diffuse scattering was isolated from the remainder by removing the nuclear Bragg features, then subtracting a background to yield zero scattering below $Q=0.4$~\AA$^{-1}$ \cite{suppmat} . The validity of the subtraction was verified by comparing with polarized neutron diffraction data on the $x=1$ composition at the same temperature. The reverse Monte Carlo (RMC) analysis of the magnetic diffuse scattering was performed using the SPINVERT package \cite{spinvert}.

	\begin{figure}%[htbp]
	\includegraphics[width=\linewidth]{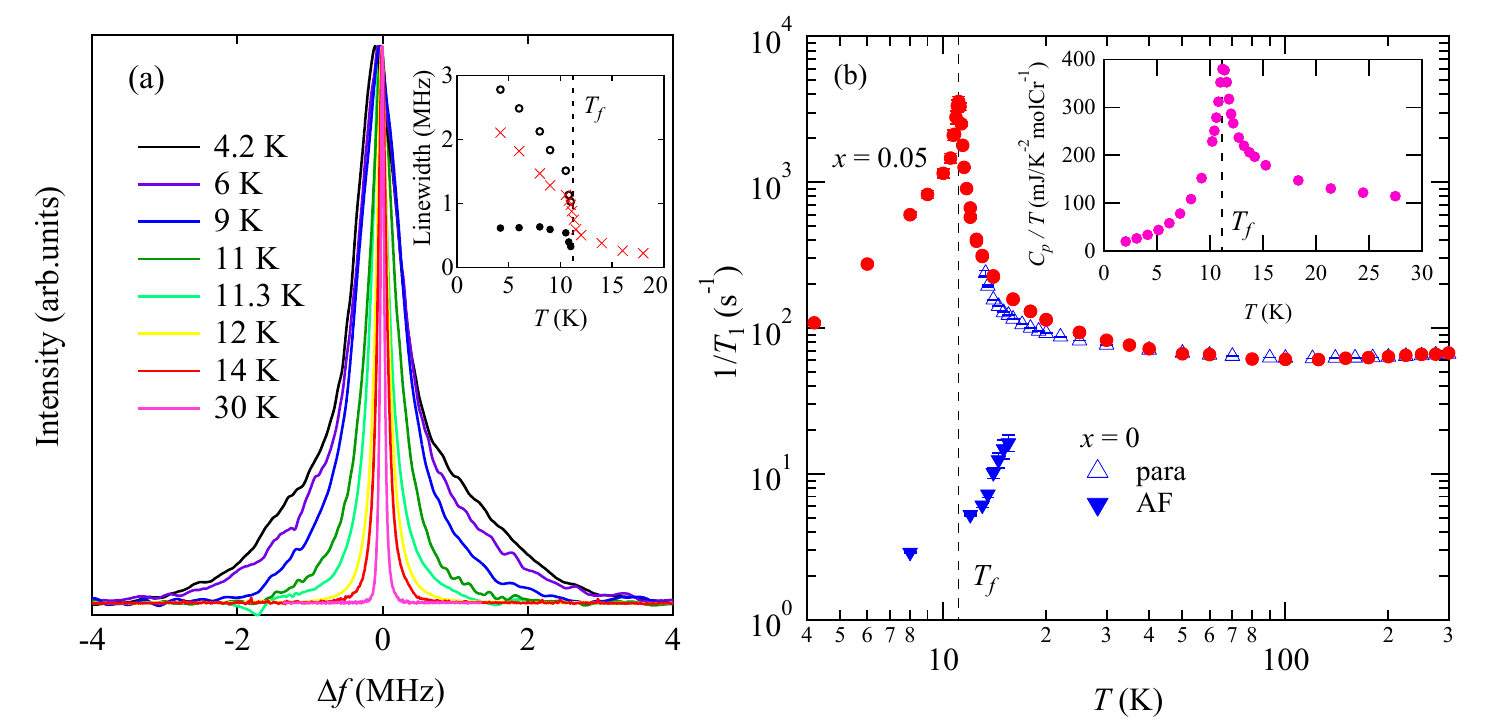}
	\caption{(a) $^{7}$Li-NMR spectra, normalized by the peak intensity. $\Delta f = 0$ corresponds to the center of gravity of the spectrum. 
The inset shows temperature dependence of the averaged linewidth of whole spectra (red crosses), the FWHM of the narrow component (closed circles), and the FWHM of the broad ones (open circles)
 (b) The temperature dependence of $1/T_1$ for the $x = 0.05$ composition (solid circles), paramagnetic (open triangles) and AF (solid triangles) components for the pure composition, respectively. The inset shows temperature dependence of the specific heat divided by temperature $C_p/T$ of the $x = 0.05$ composition. }
\label{FIG1}
\end{figure}

	The temperature dependence of the $^7$Li-NMR spectrum is shown in Fig.~\ref{FIG1}(a). In the paramagnetic state, it consists of a sharp single line without quadrupole structure, similar to the $A^\prime$=Ga$^{3+}$ compound. Below 11~K, however, the spectrum shows a marked broadening, which indicates the development of a static internal field at the $^7$Li site as a result of spin freezing. The spectra in the low-temperature phase consist of two components: a relatively narrow line whose width saturates below 9~K, and a broader one which broadens further with decreasing temperature. The temperature dependence of each line-width extracted by a double Gaussian fit are shown in the inset of Fig.~\ref{FIG1}(a). The intensity ratio of the sharp and broad components of spectra is estimated to be $1:4.7$ at $4.2$~K

	Figure~\ref{FIG1}(b) shows the temperature dependence of $1/T_1$ for both the $x=0.05$ (red circles) and $x=0$ (blue triangles) compositions \cite{beta}. In the case of the former, $1/T_1$ exhibits a sharp peak at $T_f=$11.08(5)~K, indicating critical slowing down associated with a bulk second-order magnetic transition. This transition is also evidenced by a sharp anomaly at 11.29(3)~K in specific heat, also indicative of a second-order transition [Fig.~\ref{FIG1}(b), inset]. These behaviors are in contrast with the $x=0$ compound, which shows two first-order magnetic transitions, and phase separation [Fig.~\ref{FIG1}(b)].

	Despite the clear hallmarks of a second-order transition in both specific heat and $^7$Li NMR, both SXRD and ND surprisingly indicate an absence of structural symmetry breaking below $T_f$, unlike all other chromate spinels with well-defined phase transitions [Fig. 2(a)]. Should the transition corresponds to a purely magnetic long range ordering, the only antiferromagnetic structure compatible with a cubic structural symmetry is the so-called all-in all-out structure, where the spins lie along the local $\langle 111 \rangle$ axes. On the other hand, this structure would imply zero internal field at the $^7$Li site, in contradiction with the large width of the NMR line. Neither do any sharp features consistent with all-in all-out order appear in ND.

\begin{figure}
	\includegraphics[width=\linewidth]{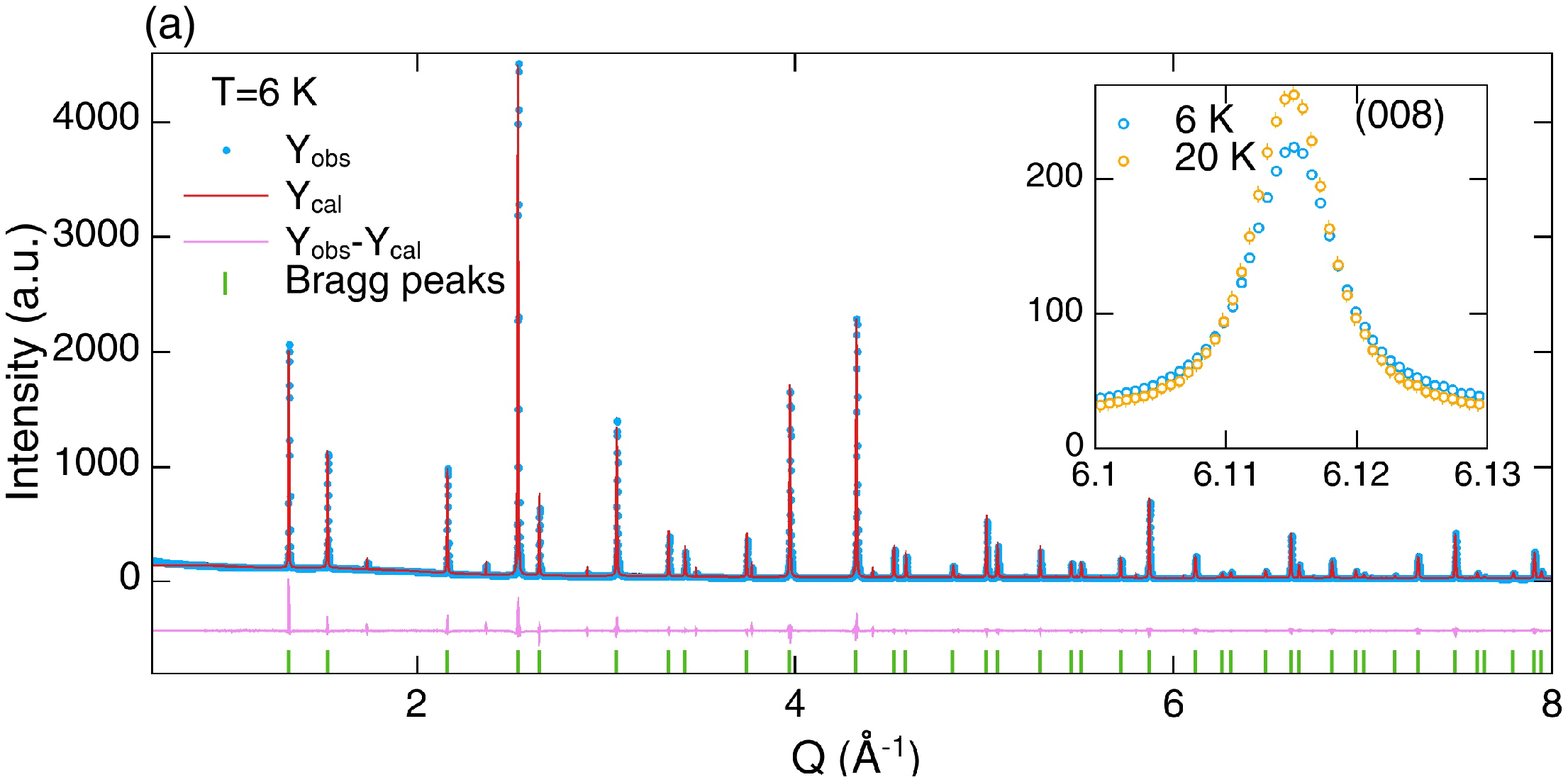} \\
	\includegraphics[width=\linewidth]{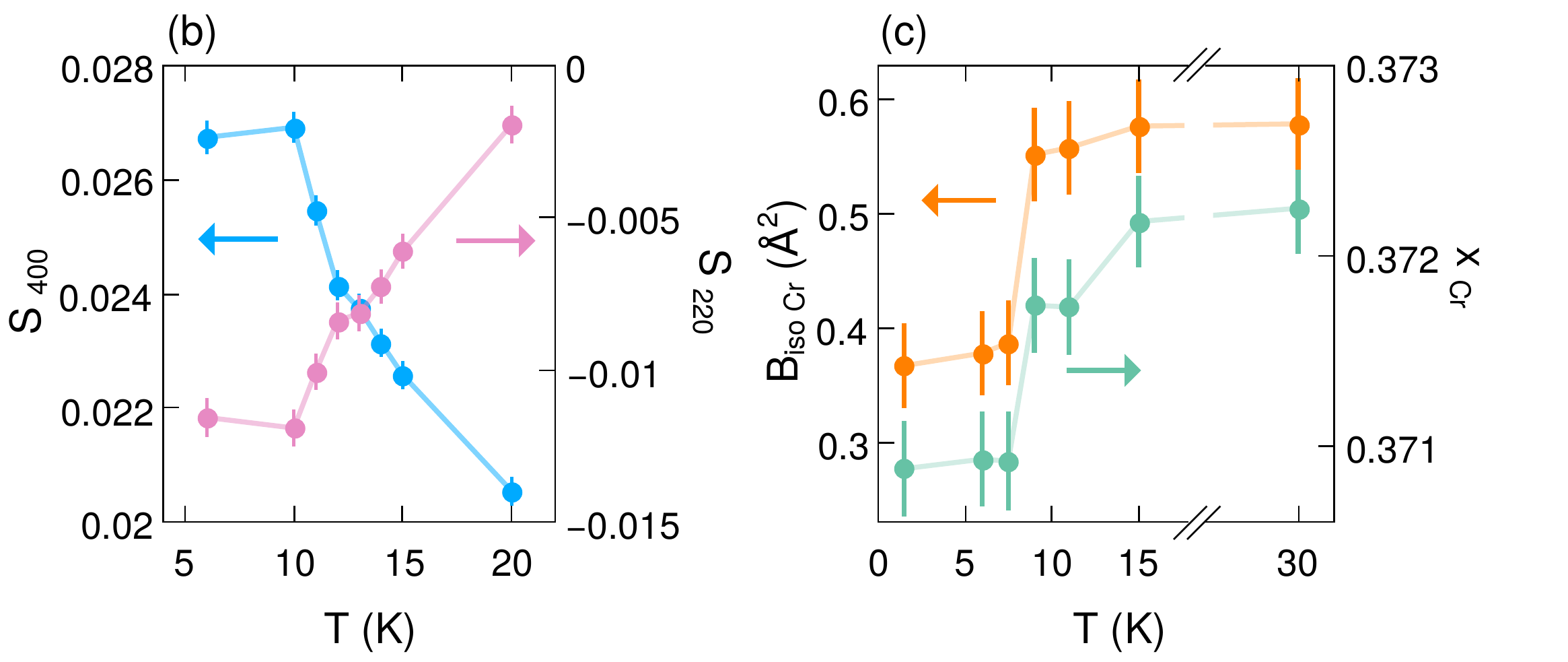} 
	\caption{(a) Synchrotron X-ray pattern at low and high $T$ with its Rietveld refinement($R_p$=12.2 $R_{wp}$=15.6 $R_e$ =4.05), (inset) T-dependence of $(008)$ peak. Peaks visible in the measured pattern and not signed with green markers come from a $\sim1\%$ Cr$_2$O$_3$ impurity. Pattern refined with resulting fit parameters: Temperature dependence of (b) strain parameters $S_{400}$ and $S_{220}$, extracted from SXRD, and (c) $B_{iso}$ and Cr fractional position parameter $x$ obtained from ND.}
\label{FIG2}
\end{figure}	
	
	While no peak splittings are observed on crossing $T_f$, some broadening of Bragg peaks with indices $(h00)$ and $(hk0)$ is observed on cooling towards $T_f$. This implies that the local symmetry is tetragonal, as expected from magneto-elastic coupling within the bilinear-biquadratic model \cite{tchernyshyov02}. To identify the changes in the structure on crossing $T_f$, we plot the temperature dependence of the $S_{400}$ and $S_{220}$ strain broadening parameters \cite{shkl} corresponding to these families of peaks, as well as the fractional Cr$^{3+}$ position parameter and isotropic displacement parameter in Fig. \ref{FIG2}(b,c). All parameters are found to evolve continuously before freezing at $T_f$, with the latter doing so in a step-like fashion. This emphasizes the strong magneto-elastic coupling in the system. Furthermore, the apparent freezing of all parameters below $T_f$ is consistent with the spin freezing observed in NMR.

\begin{figure}[b]
    \includegraphics[width=\linewidth]{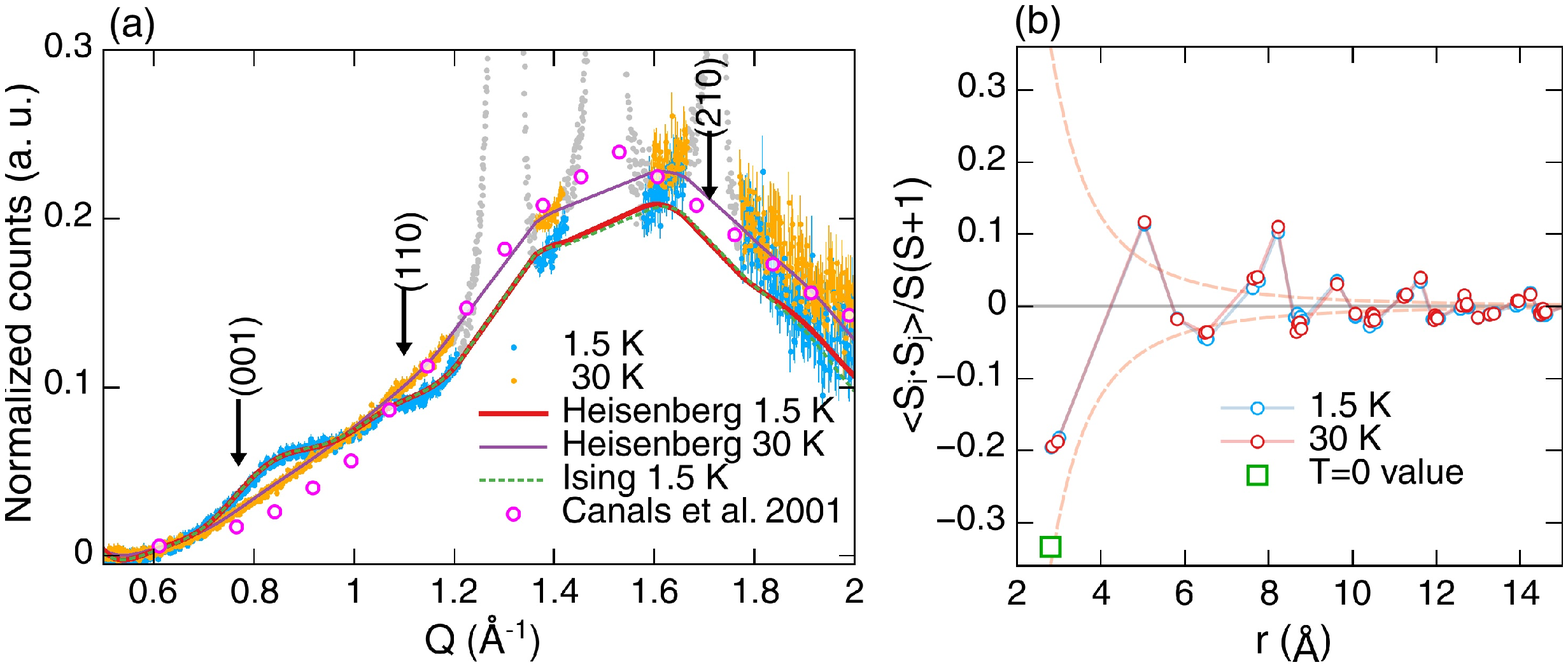} \\
        \includegraphics[width=\linewidth]{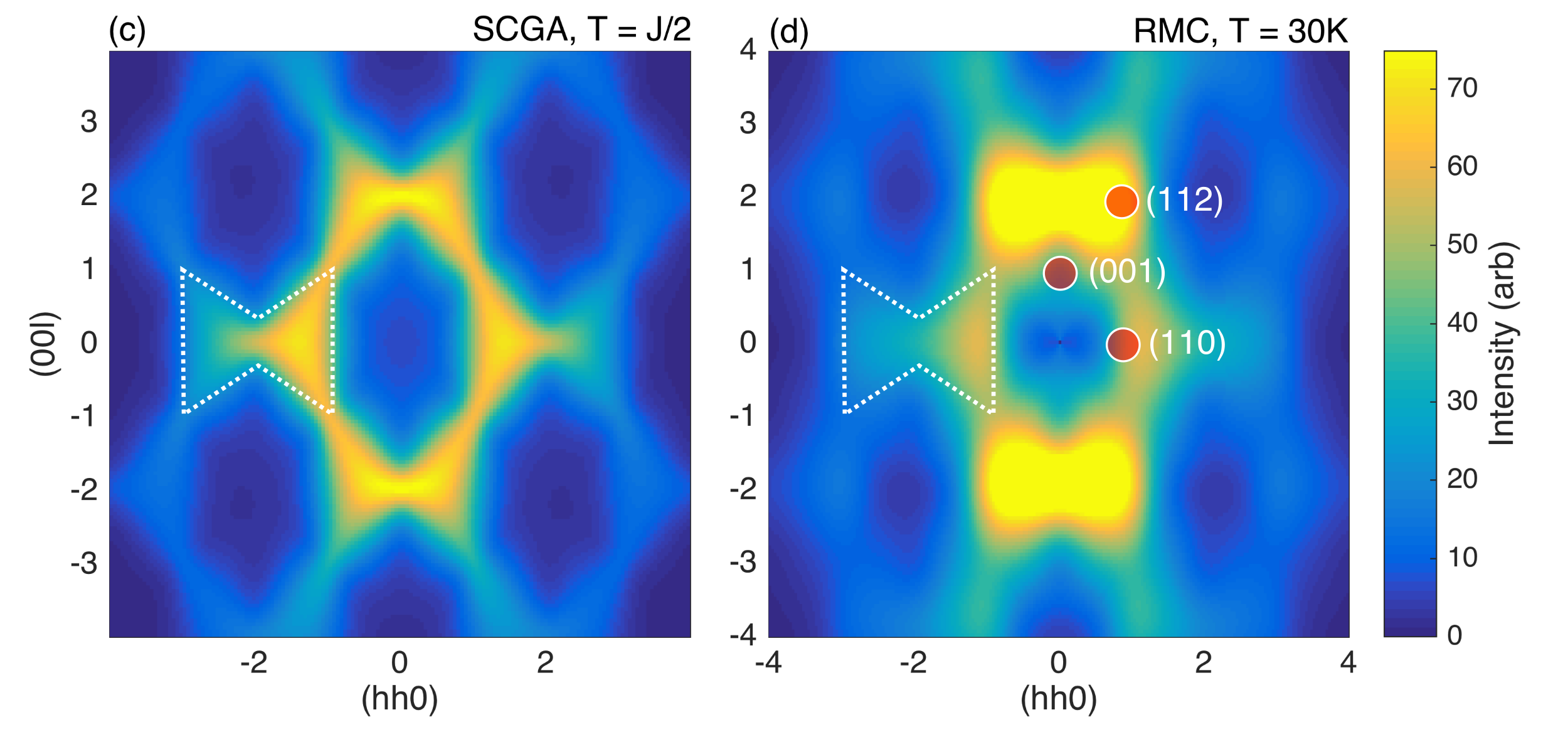}
	\caption{(a) Diffuse scattering at 1.5 K and 30 K with RMC fits using free (Heisenberg) and constrained (Ising) spins parallel to (001) for the former. (b) Real space spin-spin correlation functions $\langle S_0\cdot S_i\rangle/S(S+1)$ versus $r$ for both temperatures. The dashed lines indicate the expected envelope for dipolar correlations $\propto 1/r^3$, and the green square the expectation for  $\langle S_0\cdot S_1\rangle/S(S+1)$ as $T\rightarrow0$. (d) Reconstructed single crystal scattering from the 30 K Heisenberg fit compared with (c) ``bow-tie'' scattering (highlighted by the white dashed polygon) calculated for the ``breathing'' pyrochlore antiferromagnet within the self-consistent Gaussian approximation (SCGA), for $B_f=0.6$ at $T/J=0.5$ \cite{benton15}.}
	\label{FIG3}
\end{figure}

	Turning to the magnetic diffuse scattering, the data at $30$~K, well above $T_f$, show a broad, diffuse feature with maximum centred around $Q=1.55$~\AA$^{-1}$ [Fig.~\ref{FIG3}(a)]. This is compatible with expectations for the undistorted pyrochlore lattice, and implies the presence of Coulomb-phase-like power law spin-spin correlations \cite{isakov04}. Cooling to $15$~K, a weak, but sharp peak is observed at the $(110)$ position, as for $x=0$. Because this appears above $T_f$ and appears to be temperature independent on further cooling, it is ascribed to the presence of a small amount of $x=0$ phase in the sample. The only intrinsic changes in the magnetic scattering on crossing the transition are therefore a slight redistribution of the diffuse scattering towards the positions $Q$ = 0.8 \AA$^{-1}$ [near $(100)$], $1.1$~ \AA$^{-1}$ $(110)$, 1.73 \AA$^{-1}$ $(210)$, and 1.87 \AA$^{-1}$ $(112)$ [Fig.~\ref{FIG3}(a)].

	To understand the apparent paradox of the presence of a phase transition, on the one hand, and the absence of any peak splittings or (intrinsic) magnetic peaks, on the other, we investigate the changes in the real space spin-spin correlations by performing RMC fits of the magnetic diffuse scattering \cite{paddison13}. At $30$ K, the extracted normalized real-space spin-spin correlation function $\langle S_0 \cdot S_i \rangle/S(S+1)$  indicates antiferromagnetic nearest neighbour correlations, with $\langle S_0 \cdot S_1 \rangle/S(S+1) = -0.2$ [Fig.~\ref{FIG3}(b)]. This is consistent with conventional MC simulations of both the undistorted and ``breathing'' pyrochlore lattices at finite temperature, where the ground states are Coulomb liquids \cite{shannon10,benton15}. However, the correlations do not quite follow the expected dipolar form: for example, $\langle S_0 \cdot S_3 \rangle/S(S+1)$, which corresponds to the next-nearest-neighbour distance along $\langle 110 \rangle$, is negative.
	
	Reconstructing the single crystal scattering from the RMC spin configurations in the $(hhl)$ [Fig. 3(c)] and $(hk0)$ planes \cite{spinvert} provides some clues as to why this is: intensity is observed at positions consistent with the propagation vector $\mathbf{k} = (001)$, which manifests in our experiments as scattering around the Bragg positions listed previously. The scattering maps also allow us to identify ``bow-tie'' features characteristic of the underlying Coulomb liquid state \cite{benton15} [Fig. 3(c-d)]. The overall picture of $\mathbf{k} = (001)$ short range order superimposed on ``bow-tie'' features remains unchanged to 1.5~K, although the former grow slightly in intensity below $T_f$. This is also true of $\langle S_0 \cdot S_i \rangle/S(S+1)$, which is nearly indistinguishable between the high and low temperature datasets [Fig.~\ref{FIG3}(b)], emphasising that the order parameter of the transition cannot be dipolar.

	Classical MC simulations for the bilinear-biquadratic model with a Gaussian bond disorder $\Delta$ (here related to the substitution $x$) indicate a quadrupolar nematic transition to a collinear state with persistent Coulomb-like correlations for $b<0$, $\mathcal{P}=0$, and small $\Delta$ \cite{shinaoka14}. As $\Delta$ ($x$) is increased, this nematic transition becomes concurrent with spin freezing. Because of (i) the lack of magnetic Bragg peaks at $T<T_f$, despite a clear phase transition; (ii) the tendency towards collinear spin arrangements in $x=0,1$, implying $b<0$ for all $x$; (iii) the local symmetry lowering, consistent with such spin arrangements; and (iv) the spin freezing observed in NMR, we speculatively assign the phase transition to concurrent nematic order and freezing. To test this assignment, we perform further RMC simulations on the $1.5$~K data with the spins constrained to lie along any of three high symmetry directions in the cell [(001), (110), and (111)] -- \textit{i.e.} collinear spins \cite{ising}. Several directions are modelled due to the cubic symmetry of the system; while all should reproduce the experimental scattering, the corresponding spin configurations generally differ. With collinear spins, the data is modelled nearly as well as the Heisenberg case ($\chi^2_{\textnormal{Ising}}/\chi^2_{\textnormal{Heis}} \sim 1.05$) for all spin directions. 
		
	To verify that this assumption is consistent with our NMR results, we simulated the NMR spectrum corresponding to the collinear RMC spin configurations \cite{suppmat}. The main component of the simulated spectrum, ascribed to the broad part of the magnetic diffuse scattering (and hence the ``bow-tie'' features), is triangular, and reproduces the shape of the broad component in the experimental NMR spectrum at $4.2$~K [Fig.~\ref{FIG1}(a)]. On the other hand, the sharp component, corresponding to the $\mathbf{k}=(001)$ short-range order, is underestimated by our RMC simulations. This discrepancy is likely due to the difficulty of fitting $\mathbf{k}=(001)$ magnetic diffuse scattering near nuclear positions in the experimental data, the different temperatures of the NMR ($4.2$~K) and neutron ($1.5$~K) measurements, and the inhomogenous decay of the NMR intensity due to slow fluctuations \cite{suppmat}. Scaling the width of the triangular component to the NMR data, the moment size on the Cr$^{3+}$ site is estimated to be $gS=1.3$~$\mu_B$ at 4.2~K. Although this quantity is model-dependent, the reduced moment could indicate the present of fast quantum or thermal fluctuations in the ground state.

	The RMC spin configurations may be further analyzed by examining the collinear spin arrangements on individual tetrahedra: for all directions of the anisotropic axis, configurations with two spins up and two down ($uudd$) are favored over other configurations. The local constraint $\sum_{i\in\textrm{tet}}{\vec{S}_i=0}$ obeyed by these configurations is responsible for the ``bow-tie'' scattering in the reconstructed single crystal patterns. The ratio of $uudd$ to three-up one-down ($uuud$, or vice versa, $uddd$) configurations is typically in excess of 8 for the $(001)$ axis, versus $2-4$ for the other directions. Assuming that all interactions in the system are antiferromagnetic, this solution is the most energetically favorable, and we will therefore focus on it in the subsequent analysis. 
	
	Following Tchernyshyov et al. \cite{tchernyshyov02}, each $uudd$ tetrahedron may be given a color according to the arrangement of the two ferro- and four antiferromagnetic bonds on each tetrahedron, and correspondingly, the direction of the local tetragonal distortion. Ferromagnetic (long) bonds along $\langle 110 \rangle$ correspond to blue (B), $\langle 101 \rangle$ green (G), and $\langle 011 \rangle$ red (R). The refined spin structures exhibit a majority of B tetrahedra. In the case of the $x=0$ and $x=1$ compounds, the magnetic orders respectively correspond to RG and BG color arrangements on the diamond lattice formed by the tetrahedra, with the spins directed along the $c$ axis of the tetragonal cell. The observed growth of the $\mathbf{k} = (001)$ scattering may thus be associated with domains containing combinations of BR and BG. Interestingly, this means that the ground state of the present material lies closer to the $x=1$ ordered structure, despite its chemical proximity to the $x=0$ compound. The BR/BG short-range order is furthermore consistent with the negative $\langle S_0 \cdot S_3 \rangle/S(S+1)$ found in the Heisenberg fits, as well as the broadening (but lack of splitting) of the structural diffraction peaks.

\begin{figure}
	\includegraphics[width=\linewidth]{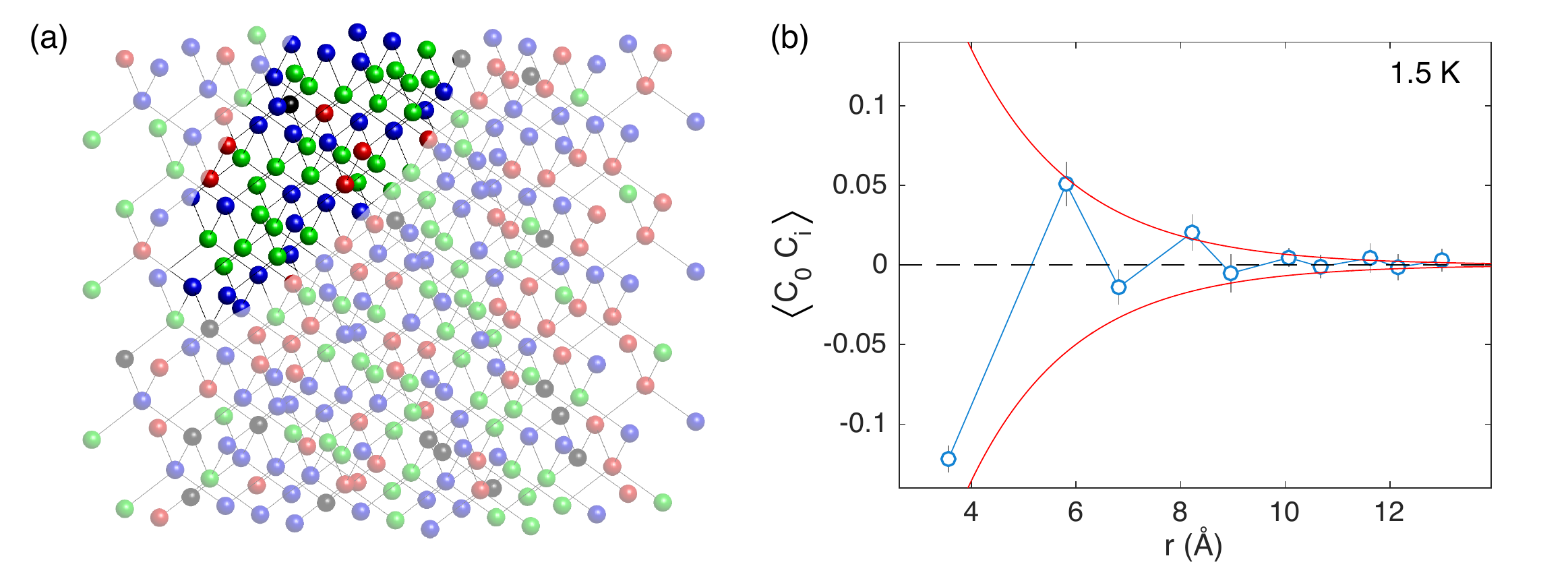} \\
	\caption{(a) Real space plot of a color RMC configuration at 1.5 K ($2\times 2\times 2$ unit cells). The black spheres indicate $uuud$ tetrahedra (or vice versa), and the highlighted area shows a region with predominantly blue-green correlations. (b) Radial dependence of color-color correlations $\langle C_0 C_i \rangle$ at $1.5$~K. The red lines correspond to the expected envelope for two-color correlations decaying exponentially with the correlation length $\xi_c = 2$~\AA.}
	\label{FIG4}
\end{figure}

	To quantify the spatial extent of the opposite color correlations, we compute the color correlation function $\langle C_0 C_i \rangle$, defined such that $\langle C_0 C_i \rangle=1$ for the same and $\langle C_0 C_i \rangle=-1$ for different colors between tetrahedra, and weighted to account for the unequal color populations. We thus find $\langle C_0 C_1 \rangle=-0.14$ and opposite-color short range order, with an exponential radial decay characterized by the correlation length $\xi_c =$2~\AA{}. While this is short, two-color (BG and BR) domains as large as 20~\AA{} can be identified by visual inspection of the color configurations [Fig \ref{FIG4}(b)]. The scattering around $(001)$ (0.8 \AA{}$^{-1}$) can be understood as resulting from disorder between the minority colors (R and G) in the two-color pattern -- in the case of perfect color order, it would vanish. In this sense, the low temperature state shows some commonalities with the partially ordered phase predicted from MC simulations of the pyrochlore lattice with further neighbor couplings in \cite{chern08} (this state is also found in the three-state Potts model \cite{lapinskas98}). The partially ordered phase has a single color on all up-pointing tetrahedra, and a distribution of other colors on down-pointing tetrahedra.

		There are, however, some uncertainties concerning the above interpretation. Most importantly, the existing MC simulations only consider the undistorted pyrochlore lattice, and it is not known how the phase diagram changes when the ``breathing'' distortion is introduced. Indeed, in the undistorted case, the transition is expected to be weakly first-order for small $x$, only becoming second-order deep inside the spin glass regime, where the nematic transition coincides with spin freezing. Both of these statements contradict our observation of a second-order transition and spin freezing at small $x$, as well as the appearance of a conventional spin glass phase at larger $x$ \cite{okamoto15}. On the other hand, the theoretically predicted spin glass state at large $x$ is expected to be exceptionally robust towards magnetic field, like the present low-temperature phase \cite{suppmat} -- similar behaviour is observed below $T_f$ in pure Y$_2$Mo$_2$O$_7$, where no clear nematic transition is observed. It is finally not evident why the nematic transition occurs at all, given the presence of the further neighbor couplings which presumably play a role in the ordering of the $x=0$ and $x=1$ compounds. These points will hopefully be clarified by future MC simulations and experimental work.

	To conclude, we have shown that LiGa$_{0.95}$In$_{0.05}$Cr$_4$O$_8$ undergoes a single, apparently second-order, transition at $T_f$ = 11 K. This transition corresponds neither to magnetic long range order nor a structural symmetry breaking, but is rather ascribed to nematic (collinear) spin freezing. Upon cooling, correlations corresponding to the propagation vector $\mathbf{k} = (001)$ are enhanced. Assuming that the spins lie along $(001)$, these correspond to short-range order between collinear spin configurations on the tetrahedra. The transition thus shares features with both the nematic and partial ordering transitions anticipated for the pyrochlore Heisenberg model with perturbations.
	
\begin{acknowledgments}
We gratefully acknowledge T. Fennell for his careful reading of and useful comments on the manuscript. We also thank N. Shannon and O. Benton for sharing their SCGA data, and S. Hayashida and J. A. M. Paddison for useful discussions. This work was supported by JSPS KAKENHI (Grant Nos. 25287083 and 16J01077). Y. T. was supported by the JSPS through the Program for Leading Graduate Schools (MERIT).
\end{acknowledgments}
	
% Create the reference section using BibTeX:
\bibliographystyle{apsrev}
%\bibliography{draft_1Notes}
\bibliography{draft_6_shortened_arxivNotes}

\begin{thebibliography}{36}
\bibitem[{mag()}]{magstr}
\bibinfo{note}{We believe this to be the magnetic space group $C_A222_1$,
  belonging to the $mM_5$ irreducible representation [Miller-Love notation,
  parent space group $I\protect \mathaccentV {bar}016{4}m2$, $\protect \mathbf
  {k}=(001)$], rather than $P_I\protect \mathaccentV {bar}016{4}m2$ ($mM_1$)
  \cite{reportpsi}, as proposed in ref \cite {saha16}.}
\bibitem[{bet()}]{beta}
\bibinfo{note}{The temperature dependence of $\beta$ is shown in
  \cite{suppmat}.}
\bibitem[{shk()}]{shkl}
\bibinfo{note}{$S_{hkl}$ is proportional to the variance of the interplanar
  distance \cite{stephens99}.}
\bibitem[{isi()}]{ising}
\bibinfo{note}{We note that the assumption of collinear spins cannot be
  justified on the basis of a finite quadrupolar order parameter from the
  high-temperature Heisenberg fits, and that the Ising fits do not converge at
  high temperature.}
\end{thebibliography}

%
% ****** End of file template.aps ******

\clearpage

\renewcommand\thefigure{S\arabic{figure}}  

\section*{SUPPLEMENTAL MATERIAL}

\setcounter{figure}{0}   

\subsection*{Nuclear spin-lattice relaxation rate 1/$T_1$}
\begin{figure}[htbp]
	\includegraphics[width=0.9\linewidth]{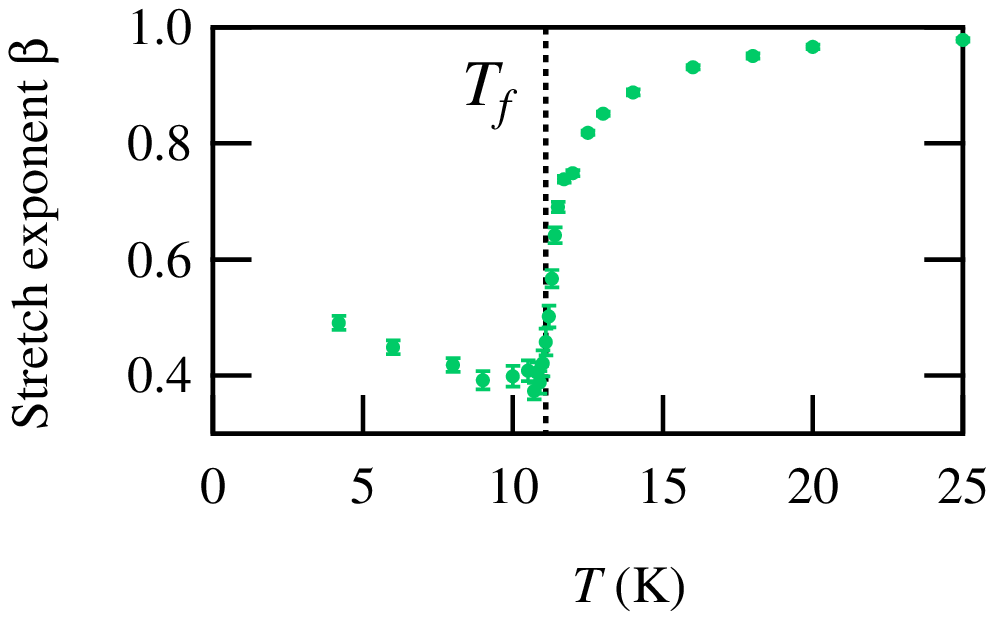}%
	\caption{Temperature dependence of stretch exponent $\beta$ for  $^7$Li nuclei in a field of 2~T. }
	\label{Beta}
	\end{figure}

The nuclear spin-lattice relaxation rate 1/$T_1$ was determined by fitting the recovery curves of the spin-echo intensity at the spectral peak after an inversion pulse as a function of time $t$ to the stretched-exponential function:
\begin{equation}
 I(t)=I_{\mathrm{eq}}-I_0 \exp \left[-\left(t/T_1\right)^{\beta}\right], 
\end{equation}
where $I_{\mathrm{eq}}$ is the intensity at the thermal equilibrium and $\beta$ is a stretch exponent that provides a measure of inhomogeneous distribution of 1/$T_1$. The case of homogeneous relaxation corresponds to $\beta$ = 1.
Figure~\ref{Beta} shows the temperature dependence of $\beta$, which decreases from $1$ on cooling below 30~K and reaches a minimum around $T_f$. A stretched exponential relaxation with $\beta<1$ is observed when a distribution of $1/T_1$ arises due to inhomogeneity.

\subsection*{NMR spectra and spin-echo decay time $T_2$}

\begin{figure*}[htbp]
	\includegraphics[width=0.75\linewidth]{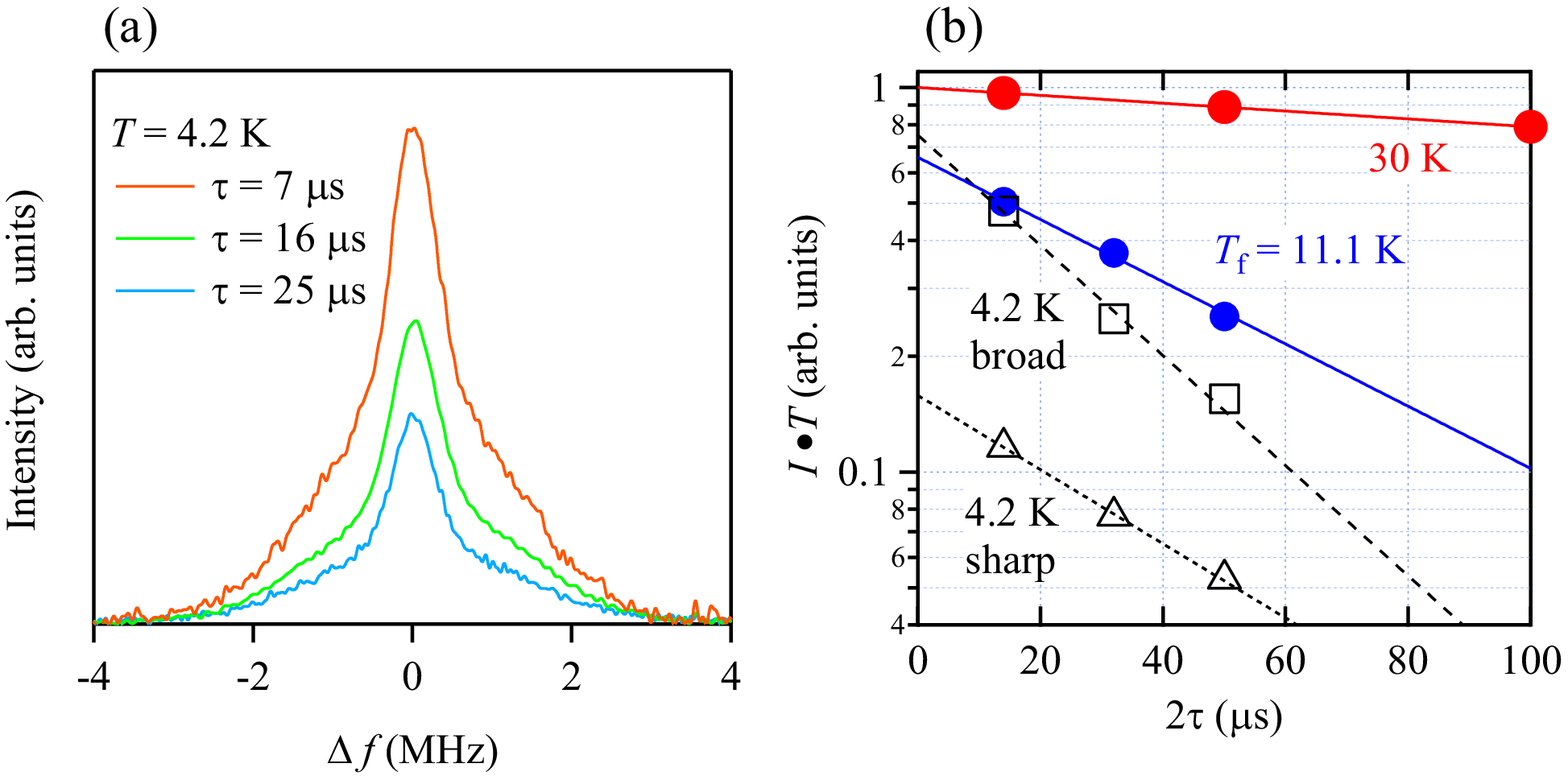}
    \caption{(a) $\tau$ dependence of spectra at 4.2~K. (b) $\tau$ dependence of the integrated intensity of NMR spectra. }
        \label{IntegInt}
\end{figure*}

In order to track the temperature-dependence of the signal intensity and spin-echo decay time $T_2$, we performed NMR spectrum measurements with different pulse separation times $\tau$ in the spin-echo pulse sequence $\pi/2$-$\tau$-$\pi$. Figure~\ref{IntegInt}(a) shows the $\tau$-dependence of NMR spectra at the base temperature. Although both sharp and broad components of the spectra have fast $T_2$, and consequently show a reduction of the intensity for longer $\tau$, the spectral shapes are almost identical within the range of the $\tau$ probed. The spectra shown in Fig.1(a) in the main paper were measured with $\tau=25$~\textmu s. 

The integrated intensities of the spectra $I$ (multiplied by $T$ to cancel out Curie law for the nuclear moments) are plotted against $2\tau$ in Fig.~\ref{IntegInt}(b); the $I$ corresponding to the sharp and broad components at base temperature were extracted by a double Gaussian fit, and are plotted separately. The fitting function for each dataset is $I(t)T=I_0T\exp(-t/T_2)$, where $T_2$ is the spin-echo decay time. The $T_2$ thus extracted are 425(1) ($30$~K$>T_f$), 54(3) ($11$~K$\sim T_f$), 45(1) ($4.2$~K, sharp), and 30(2)~\textmu s ($4.2$~K, broad). $T_2$ drops sharply at $T_f$=11.1~K and does not recover even at base temperature, as is the case for other frustrated systems with slow low-temperature dynamics.
In order to quantify the loss of NMR intensity below $T_f$, all data were normalized by the extrapolated $I(0)T$ at 30~K [Fig.~R2(b)]. The resulting $I(0)T$ are 0.66(2), 0.16(0), and 0.75(4) at 11.1~K, and for the sharp and broad component at 4.2~K, respectively. This indicates that 91~\% of the intensity is conserved at the base temperature. Finally, we note that the ratio of the intensities corresponding to the two components is estimated to be  $1:4.7$ at the base temperature.

\subsection*{Neutron and X-ray diffraction}
The sample used in the synchrotron X-ray powder diffraction experiment was enclosed in a silica capillary of $3$ mm diameter, and loaded into a $^4$He flow cryostat with a base temperature of 5 K. Neutron powder diffraction patterns were measured on 5.8 g of powder loaded in an $8$ mm diameter vanadium can. 

\begin{figure*}[htbp]
	\includegraphics[width=0.8\linewidth]{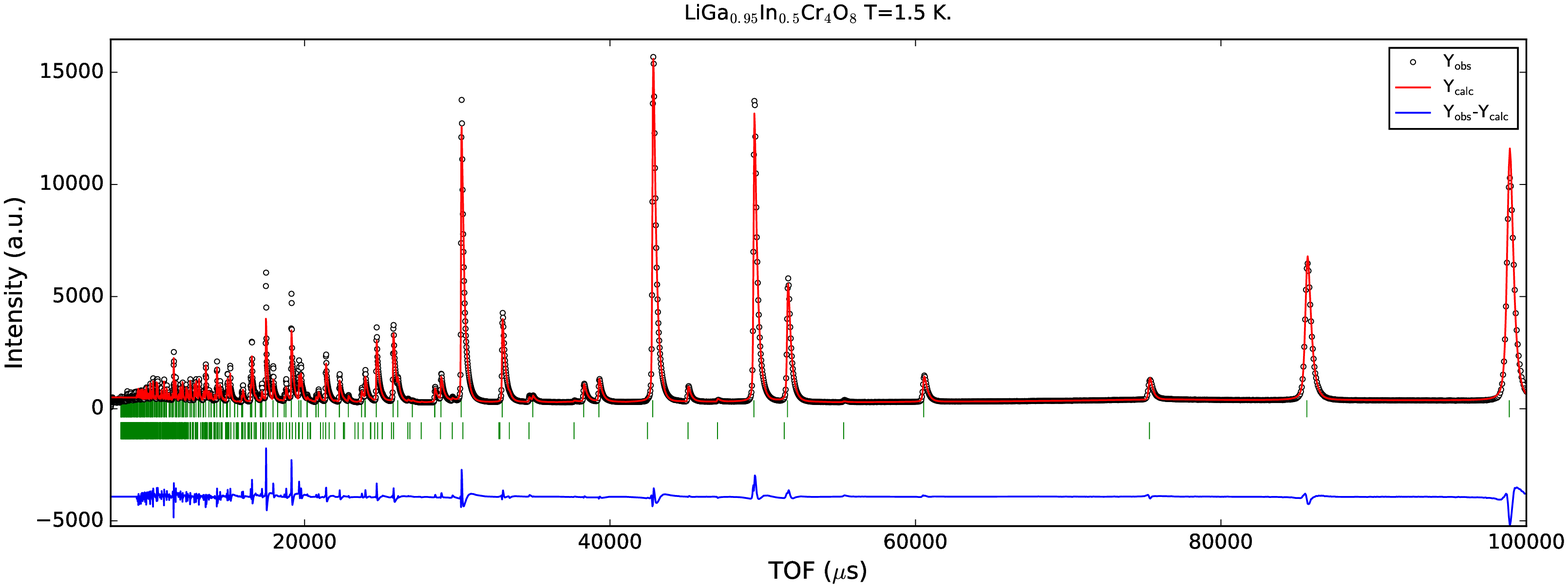}%
	\caption{Measured time-of-flight neutron diffraction pattern (black points) at $1.5$~K and its refinement (red line). The data was measured on banks 5 and 6 of the WISH diffractometer in backscattering geometry.}
	\label{NDpattern}
	\end{figure*}

The measured diffraction patterns were analysed by means of Rietveld refinement. Part of one refined neutron diffraction pattern is presented in Fig.~\ref{NDpattern}, and parameters showing the goodness of fit for all detector banks and temperatures are presented in Tab.~\ref{fit_params}. Due to the fact that the peak profile of WISH is difficult to describe with the commonly used back-to-back exponential function and high-$Q$ parts of the patterns suffer from significant peak overlap, we present the weighted $R$ parameters for the Rietveld refinement and LeBail profile matching (representing the best possible fit for the given set of profile parameters) along with the expected $R$ value based on the data set's statistics. The strain ($S_{hkl}$) parameters, whose temperature dependence is presented in main body of this work, were extracted from the SXRPD data, due to the superior resolution of these measurements. On the other hand, $B_{iso}$ and the fractional $x$ coordinate of chromium were derived from refinements of neutron powder diffraction data. The refined structural parameters are shown in Table~\ref{Rietv_params}, where position values are retrieved from SXRD patterns and $B_{iso}$ parameters come from NPD data.

		\begin{table*}
	\caption{Parameters showing the goodness of fit for high angle detector banks of WISH. Where $R_{Rwp}$ and $R_{Lwp}$ are weighted $R$ paraneters for Rietveld and LeBail refinements respectively and $R_{exp}$ is $R$ expected.\label{fit_params}}
	\begin{ruledtabular}
	\begin{tabular}[c]{cccc}
    $T$ (K)	& Banks 5 \& 6 									& Banks 4 \& 7								& Banks 3 \& 8								\\ \hline
    1.5		&  $R_{Rwp}$=7.11 $R_{Lwp}$=6.32 $R_{exp}$=0.64		& $R_{Rwp}$=6.86 $R_{Lwp}$=6.13 $R_{exp}$=0.65	& $R_{Rwp}$=6.32 $R_{Lwp}$=5.81 $R_{exp}$=0.67	\\
    6		&  $R_{Rwp}$=7.15 $R_{Lwp}$=6.37 $R_{exp}$=0.66		& $R_{Rwp}$=6.88 $R_{Lwp}$=6.15 $R_{exp}$=0.68	& $R_{Rwp}$=6.34 $R_{Lwp}$=5.83 $R_{exp}$=0.69	\\
    11		&  $R_{Rwp}$=7.98 $R_{Lwp}$=7.12 $R_{exp}$=0.57		& $R_{Rwp}$=7.53 $R_{Lwp}$=7.16 $R_{exp}$=0.58	& $R_{Rwp}$=6.85 $R_{Lwp}$=6.80 $R_{exp}$=0.60	\\
    15		&  $R_{Rwp}$=7.87 $R_{Lwp}$=7.09 $R_{exp}$=0.56		& $R_{Rwp}$=7.75 $R_{Lwp}$=7.24 $R_{exp}$=0.40	& $R_{Rwp}$=7.37 $R_{Lwp}$=6.98 $R_{exp}$=0.41	\\
    30		&  $R_{Rwp}$=7.96 $R_{Lwp}$=7.19 $R_{exp}$=0.39		& $R_{Rwp}$=7.88 $R_{Lwp}$=7.39 $R_{exp}$=0.40	& $R_{Rwp}$=7.51 $R_{Lwp}$=7.10 $R_{exp}$=0.41	\\
	\end{tabular}
	\end{ruledtabular}
  \end{table*}
	
	\begin{table*}
	\caption{Structural parameters obtained by refinement of neutron and synchrotron X-ray powder diffraction data at 6 K.\label{Rietv_params}}
	\begin{ruledtabular}
	\begin{tabular}[c]{ccccccc}
    Atom	& Wyckoff position	& $x$			& $y$			& $z$			& Occupancy	& $B_{iso}$ (\AA$^2$)	\\ \hline
    Li1	& $4a$				& 0				& 0				& 0				& 0.994(2)	& 2.175(171)				\\
    Ga1	& $4d$				& 0.75			& 0.75			& 0.75			& 0.932(6)	 & 0.406( 60)				\\
    Ga2	& $4a$				&  0				& 0				& 0				& 0.006(4)	& 0.406( 60)				\\
    Li2	& $4d$				& 0.75			& 0.75			& 0.75				& 0.002(4)	& 2.175(171)				\\
    In	& $4d$				& 0.75			& 0.75			& 0.75			& 0.066(6)	& 0.406( 60)				\\
    Cr	& $16e$				& 0.37185(4)		& 0.37185(4)		& 0.37185(4)		& 4			& 0.385( 36)				\\
    O1	& $16e$				& 0.13702(14)	& 0.13702(14)	& 0.13702(14)	& 4			& 0.509( 20)				\\
    O2	& $16e$				& 0.61802(16)	& 0.61802(16)	& 0.61802(16)	& 4			& 0.509( 20)				\\
	\end{tabular}
	\end{ruledtabular}
  \end{table*}

\subsection*{Magnetic diffuse scattering and reverse Monte Carlo refinement}

To extract the diffuse magnetic scattering from the neutron powder diffraction data, a flat background $C$ ($C\simeq 0.3$) was subtracted from the data such that the mean intensity at $Q<0.4$ \AA$^{-1}$ was zero. This choice was justified by the fact that the high temperature diffuse scattering in the first Brillouin zone is zero (indeed, polarized neutron scattering on the $x=1$ compound, where the high temperature scattering is similar, also suggests this is the case \cite{nilsen15}), and that the WISH instrumental background is flat when using a V can and radial oscillating collimator. The nuclear Bragg peaks were removed from the ranges where the build-up of diffuse scattering was observed. RMC refinements were performed on boxes of spins containing $6\times 6\times 6$ unit cells at low temperature and $3\times 3\times 3$ unit cells at $30$~K, and the simulation was repeated 10 times for every temperature point to ensure a stable solution. The optimal value of the Monte Carlo weight parameter \cite{paddison13} for all temperature datasets was determined to be $W=0.6$ for the low-$T$ Ising-type collinear spin models and $W=1$ for the high-$T$ Heisenberg-type spin models.

The analysis of the RMC-refined spin configurations consisted of several steps. Firstly, for each simulation box, the normalized real space spin-spin correlation function was evaluated:
\begin{widetext}
\begin{equation}
\langle S_0 \cdot S_i\rangle/S(S+1)=\langle S(0) \cdot S(r)\rangle/S(S+1)=\frac{1}{n(r)S(S+1)}\sum^N_j\sum^{Z_{jk}(r)}_k\mathbf{S}_j(0)\cdot \mathbf{S}_k(r),
\end{equation}
\end{widetext}
where $\mathbf{S}_i$ is the vector of the $i$-th spin in the simulation box, $N$ is the total number of sites inside the box, and $Z_{ij}(r)$ is the number of spins in the coordination shell at distance $r$. This was done using the SPINCORREL application in the SPINVERT suite \cite{paddison13}. Despite the noticeable differences between the form of the diffuse scattering at high and low temperatures, it is difficult to distinguish the differences between the real space spin-spin correlations corresponding to the fits of those datasets. This is most likely due to the dominant contribution of the broad Coulomb liquid-like component to both patterns. Following evaluation of $\langle S_0 \cdot S_i\rangle$, the single crystal patterns were reconstructed in the $(hk0)$ and $(hhl)$ planes using SPINDIFF, also part of the SPINVERT suite. This allowed for a more diagnostic comparison with theoretical calculations, and permitted identification of the Coulomb-like and short-range ($\mathbf{k}=(001)$) ordered components of the scattering.

The third step, the calculation of color populations and correlations, applied only to the low-temperature collinear spin configurations. In the case of a single tetrahedron, the bilinear-biquadratic model yields six possible magneto-structural ground state configurations, three of which are independent with respect to a global rotation of spins. These may be given a color $c=\{R,G,B\}$ according to the direction of the ferromagnetic bond (which corresponds to the long bond in the tetragonally distorted tetrahedron), as explained in the main text and Ref. \cite{tchernyshyov02}. In the pure bilinear-biquadratic model on the pyrochlore lattice, the colors in the low-temperature nematic state are uncorrelated, whereas in the bilinear-biquadratic model with long-range couplings, they are fully ordered. By calculating the correlations between the colors, we may thus effectively separate the correlations due to the nematic and short-range ordered components identified in the second step. Since the diamond lattice formed by the tetrahedra is bipartite, a color correlation function (order parameter) which distinguishes between same (\textit{e.g.} all R) and different color (\textit{e.g.} RG) orders is considered sufficient:
\begin{equation}
%\langle C_-C_i\rangle =N^{-1}\sum_{c,c^\prime}[\delta_{c,c^\prime}(1-\bar{p}_{sd})+\bar{p}_{sd}],
\langle C_0C_i\rangle=\langle C(0)C(r)\rangle=\frac{N_s^b(r)-\bar{p}_{sd}N_d^b(r)}{Z_{tot}^b}
\end{equation}
where $N_s^b(r)$ and $N_d^b(r)$ are the number of bonds of the same and different color at distance $r$, respectively, $Z_{tot}^b$ is the total number of bonds for that distance, and 
\begin{equation}
\bar{p}_{sd}=\frac{\sum_{c=c\prime}(N_c^t)^2/N_{tot}^2}{\sum_{c\neq c\prime}N_c^tN_{c^\prime}^t/N_{tot}^2}
\end{equation}
is the mean ratio of the probabilities of finding the same color tetrahedron to a different color tetrahedron on a particular bond. The final term is required as the populations of the colors in the simulation box are generally not equal. This correlation function generates $\langle C_0C_i\rangle=1$ for all shells for a same color order, and an alternation between $\pm1$ for a different color order.

%where the sum runs over all pairs of colors $c$ and $c^\prime$, $N$ is the total number of tetrahedra, $\delta_{c,c^\prime}$ is the Kronecker delta, $\bar{p}_{sd}=\sum_{i}w_c\frac{N_c(r)}{N_{c}-N_{c}}(r)$ is the ratio of the probabilities of finding a same and different color combinations between a tetrahedron at the origin and at distance $r$, $N_i$ is the number of colour population among the terahedra and $w_i=\frac{N_i}{N_{tetr}}$ is weighting factor to account for the unequal populations of the three colors.

\subsection*{Spectrum simulation from RMC spin map}

\begin{figure}[htbp]
	\includegraphics[width=0.8\linewidth]{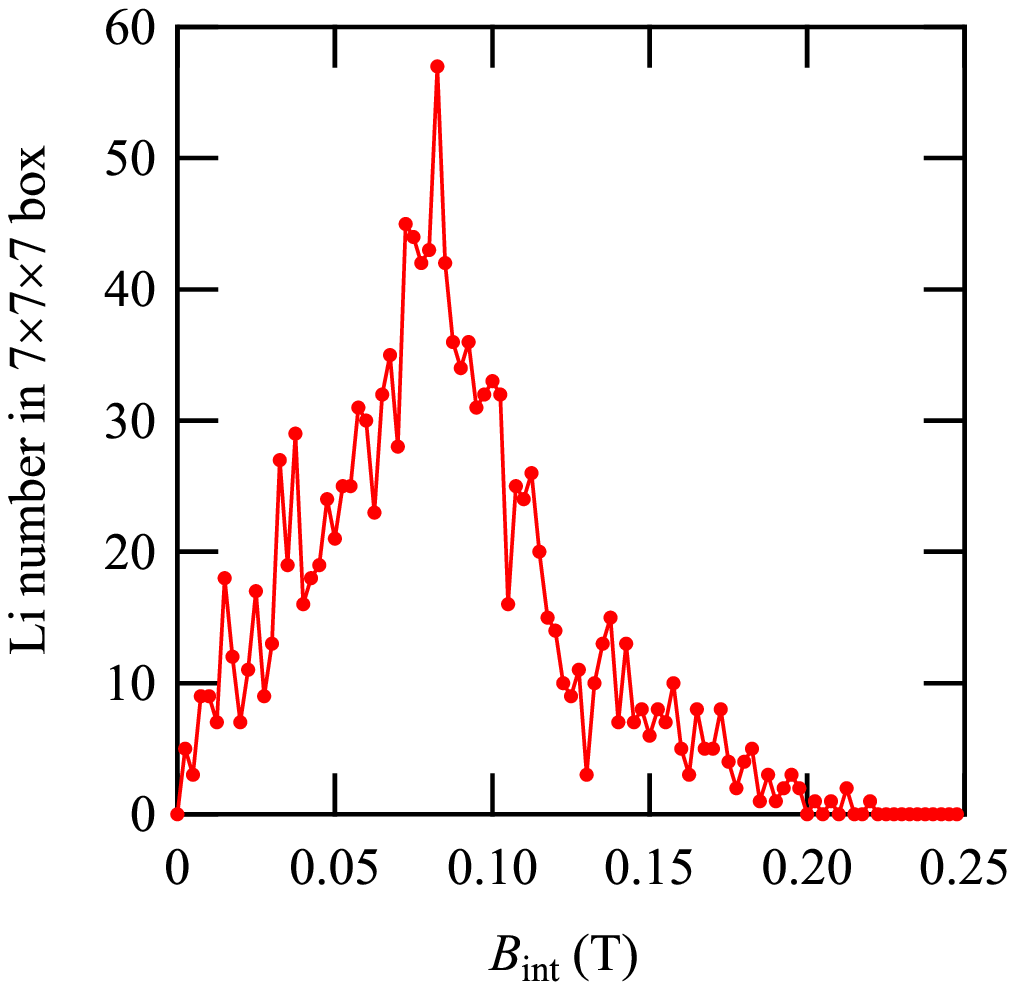}%
	\caption{Histogram of magnitude of the internal field at Li sites. The size of ordered moment in the RMC spin map is taken as 1~$\mu_{B}$ per Cr$^{3+}$ ion.}
	\label{Bhist}
	\end{figure}

The $^7$Li NMR spectrum of the low temperature phase was simulated from the spin configurations obtained by RMC. The shape of the powder averaged NMR spectra of magnetic substances is mainly determined by magnitude of the internal field at the nuclear positions. To obtain an internal field distribution at each $^7$Li site from a given spin map, we summed up the classical dipole field and transfer hyperfine field from Cr$^{3+}$ spins within a 100~{\AA} radius and from nearest-neighbor Cr$^{3+}$ sites, respectively. The transfer hyperfine coupling constant is 0.10~T/$\mu_{B}$ from the 12 nearest-neighbor Cr$^{3+}$ spins, as estimated from a $K$-$\chi $ plot above 100~K. Figure~\ref{Bhist} shows a histogram of magnitude of the internal field $B_\mathrm{int}$ calculated from the spin map with ordered moments of 1~$\mu_{B}$ per Cr$^{3+}$ ion.

	\begin{figure}[htbp]
	\includegraphics[width=0.9\linewidth]{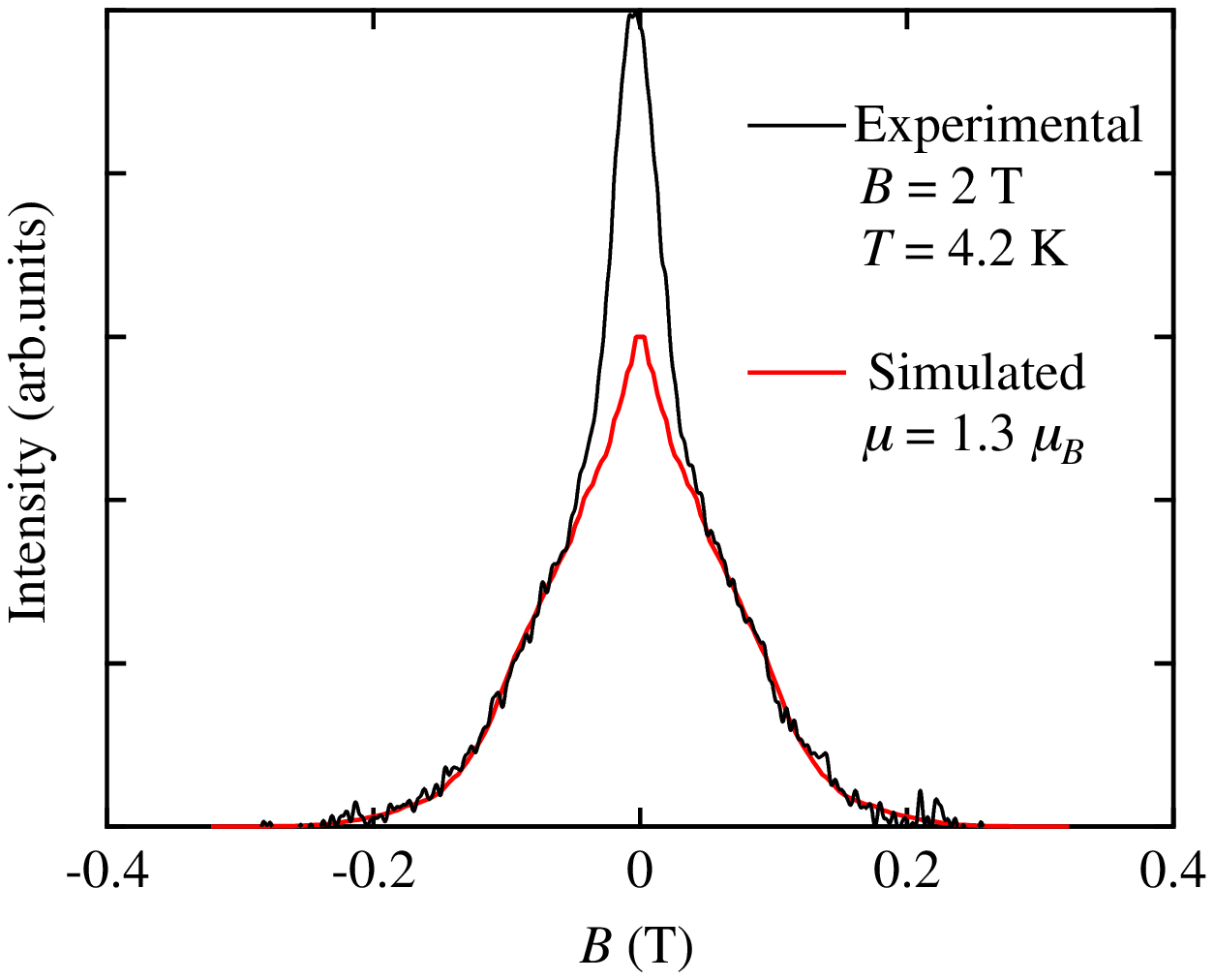}%
	\caption{Experimental NMR spectrum of the ordered phase at 4.2~K with pulse separation $\tau=7$~\textmu s(black curve) and simulated spectrum from RMC spin map (red curve). The unit of the horizontal axis of the experimental one is changed to magnetic field by dividing frequency by the gyromagnetic ratio of $^{7}$Li. The origin of the horizontal axis corresponds to the center of gravity of both spectra.}
	\label{SimulatedSp}
	\end{figure}
		
When the external field $B_\mathrm{ext}$ is sufficiently larger than $B_\mathrm{int}$, a rigid AF spin arrangement producing a single value of $B_\mathrm{int}$ yields a rectangular NMR spectrum bounded by $|B_\mathrm{int}-B_\mathrm{ext}| \leq B \leq B_\mathrm{int}+B_\mathrm{ext}$ for powder samples~\cite{yamada86} . We obtained our simulated NMR spectrum by piling up rectangle distributions whose half widths were $B_\mathrm{int}$, the horizontal axis of Fig.~\ref{Bhist}. The height of each rectangle was normalized so that its area was proportional to the vertical value of each point in Fig.~\ref{Bhist}.

The experimental NMR spectrum at 4.2~K and a simulation from an RMC spin configuration at 1.5~K are shown in Fig~\ref{SimulatedSp}; the vertical and horizontal scales of the simulated spectrum are both adjusted to match the broad component of the experimental spectrum. While the simulated spectrum appears triangular, the experimental one clearly consists of two Gaussian components with different linewidth. This discrepancy could be due to a variety of factors, including the difficulty of fitting the $\mathbf{k}=(001)$ scattering near nuclear positions, the different temperatures of the NMR ($4.2$~K) and neutron ($1.5$~K) measurements, and the inhomogenous decay of the NMR intensity due to slow fluctuations.

Based on the horizontal axis scaling factor mentioned above, the moment size of Cr$^{3+}$ spins in the wide component may be estimated to be 1.3 $\mu_B$ at 4.2~K. The moment size in the sharper component is more difficult to estimate. However, because the perfect two-color orders produce either no field distribution or a relatively narrow rectangular spectrum (the FWHM is 0.1~T at most for a full moment $gS=3~\mu_{B}$) depending on the color configurations, this contribution can readily be associated with two-color order domains.

\subsection*{Magnetic susceptibility}

\begin{figure}%[htbp]
	\includegraphics[width=\linewidth]{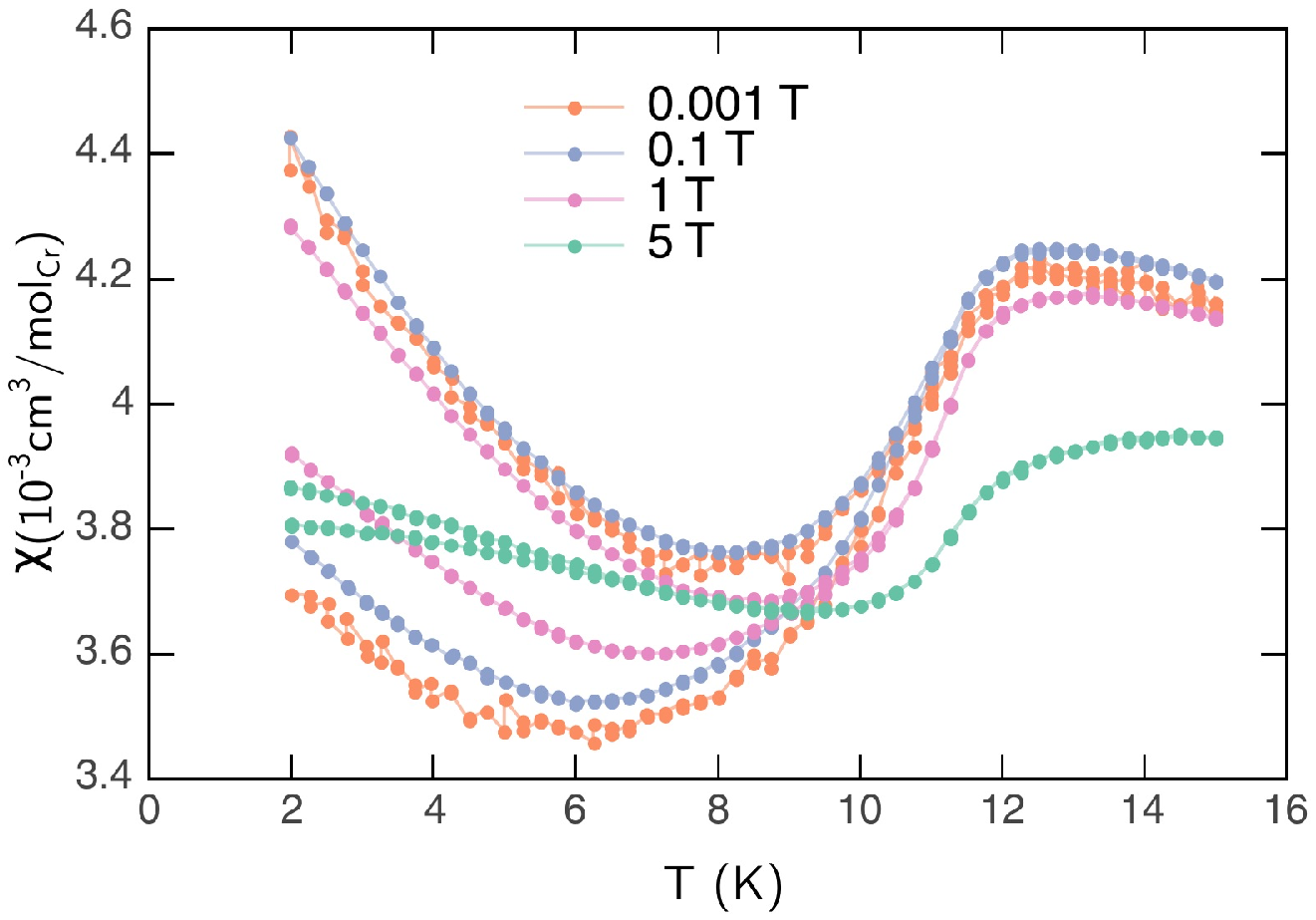}%
	\caption{Magnetic susceptibility: zero field cooled(ZFC) and field cooled(FC) measured at various values of magnetic field.}
	\label{Chi}
	\end{figure}

The zero field cooled (ZFC) and field cooled (FC) magnetic susceptibility curves for $x=0.05$ (Fig.~\ref{Chi}) show a clear splitting, indicating spin freezing below $T_f$; hence, the nematic spin freezing scenario appears likelier than the pure nematic transition. Unlike a regular spin glass, the splitting persists up to $5$~T. This robustness towards applied magnetic field is also observed in materials like SrCr$_{9p}$Ga$_{12-9p}$O$_{19}$ and Y$_2$Mo$_2$O$_7$ \cite{gardner99,silverstein14}, where the ground states are characterised by flat energy landscapes with shallow minima. Such energy landscapes, and as a result, behaviors of the ZFC and FC susceptibility, have also been predicted for the pyrochlore lattice with bilinear-biquadratic interactions and bond disorder \cite{shinaoka14}.

\end{document}